\documentclass[sigconf]{acmart}

\AtBeginDocument{%
  }
    
\usepackage{times}
\usepackage{latexsym}

\usepackage[T1]{fontenc}
\usepackage{microtype}

\usepackage[textsize=tiny]{todonotes}
\usepackage{float}
\usepackage{graphicx}
\usepackage{multirow}
\usepackage{multicol}
\usepackage{tabularx}
\usepackage{mathtools}
\usepackage{enumitem}
\usepackage{amsthm}
\usepackage{bbm}
\usepackage{comment}
\usepackage{caption}
\usepackage{subcaption}
\usepackage{wrapfig}
\usepackage{pifont}
\usepackage{tabularx}
\usepackage{booktabs}
\usepackage{url}
\usepackage{algorithm}
\usepackage{algpseudocode}
\usepackage[table,dvipsnames]{xcolor}
\usepackage{booktabs}
\usepackage{amsmath}
\usepackage{subcaption}
\usepackage{multirow}
\usepackage{soul}
\usepackage{color, colortbl}
\usepackage{wrapfig}

\usepackage{lipsum}

\usepackage[skins,breakable]{tcolorbox}
\definecolor{gray}{RGB}{236, 236, 236} 
\definecolor{lightgreen}{RGB}{32, 144, 140}  % 
\definecolor{lightorange}{RGB}{229, 231, 255} 
\definecolor{darkblue}{RGB}{15, 84, 205} % 

\setcopyright{acmlicensed}
\copyrightyear{2018}
\acmYear{2018}
\acmDOI{XXXXXXX.XXXXXXX}
\acmConference[Conference acronym 'XX]{Make sure to enter the correct
  conference title from your rights confirmation email}{June 03--05,
  2018}{Woodstock, NY}
\acmISBN{978-1-4503-XXXX-X/2018/06}

\begin{document}

\title{Learning to Route Queries to Heads for Attention-based Re-ranking with Large Language Models}

% \affiliation{%

\author{Yuxing Tian}
\affiliation{%
  \institution{Université de Montréal}
  \city{Montreal}
  \country{Canada}}
\email{yuxing.tian@umontreal.ca}

\author{Fengran Mo}
\affiliation{%
  \institution{Université de Montréal}
  \city{Montreal}
  \country{Canada}}
\email{fengran.mo@umontreal.ca}

\author{Zhiqi Huang}
\affiliation{%
  \institution{Capital One}
  \city{Boston}
  \country{USA}}
\email{zhiqi.huang@capitalone.com}

\author{Weixu Zhang}
\affiliation{%
  \institution{McGill University \& MILA}
  \city{Montreal}
  \country{Canada}}
\email{weixu.zhang@mail.mcgill.ca}

\author{Jian-Yun Nie}
\affiliation{%
  \institution{Université de Montréal}
  \city{Montreal}
  \country{Canada}}
\email{jian-yun.nie@umontreal.ca}

% \affiliation{%

% \affiliation{%

% \affiliation{%


\begin{abstract}
Large Language Models (LLMs) have recently been explored as fine-grained zero-shot re-rankers by leveraging attention signals to estimate document relevance. However, existing methods either aggregate attention signals across all heads or rely on a statically selected subset identified by heuristic rules. This solution can be suboptimal because the informative heads can vary across queries or domains. Moreover, naively combining multiple heads can degrade performance due to redundancy or conflicting ranking signals. In this paper, we propose a query-dependent head selection method, \textbf{RouteHead}, for attention-based re-ranking with LLMs. Specifically, we learn a lightweight router that can map each query to an optimal head set, and relevance scores are computed by aggregating attention signals only from these heads. Since query-to-head optimal labels are unavailable, we first construct pseudo labels via an offline search. The router represents each head with a learnable embedding and represents each query using an embedding extracted from the hidden states of the frozen LLM. Then it is trained on the pseudo labels with a sparsity regularizer. Experiments on diverse benchmarks and multiple LLM backbones show that the proposed method consistently outperforms strong baselines.
\end{abstract}
\begin{CCSXML}
<ccs2012>
   <concept>
       <concept_id>10002951.10003317.10003338.10003341</concept_id>
       <concept_desc>Information systems~Language models</concept_desc>
       <concept_significance>500</concept_significance>
       </concept>
   <concept>
       <concept_id>10002951.10003317.10003338</concept_id>
       <concept_desc>Information systems~Retrieval models and ranking</concept_desc>
       <concept_significance>500</concept_significance>
       </concept>
 </ccs2012>
\end{CCSXML}

\ccsdesc[500]{Information systems~Language models}
\ccsdesc[500]{Information systems~Retrieval models and ranking}

\keywords{Re-ranking; Large Language Model; Attention Head; Router}

\copyrightyear{2026}
\acmYear{2026}
\setcopyright{cc}
\setcctype{by}
\acmConference[SIGIR '26]{Proceedings of the 49th International ACM SIGIR Conference on Research and Development in Information Retrieval}{July 20--24, 2026}{Melbourne, VIC, Australia}
\acmBooktitle{Proceedings of the 49th International ACM SIGIR Conference on Research and Development in Information Retrieval (SIGIR '26), July 20--24, 2026, Melbourne, VIC, Australia}
\acmDOI{10.1145/3805712.3809945}
\acmISBN{979-8-4007-2599-9/2026/07}
\maketitle

\section{Introduction}

Re-ranking is a critical stage in modern retrieval systems for applications such as search and recommendation~\citep{lewis2020rag, gao2023retrieval,meng2025bridging,mo2025survey,wang2026conv}. Traditional neural re-rankers, such as bi-encoders~\citep{karpukhin-etal-2020-dense,Colbert,mo2025uniconv} and cross-encoders~\citep{nogueira2019multi, nogueira2020document,meng2024ranked}, can be effective, but they typically require substantial relevance labels and task-specific training. Recent studies show that large language models (LLMs) can perform re-ranking in a zero-shot setting by leveraging their strong ability~\citep{sachan2022improving, sun2023RankGPT, qin2024large,meng2026revisiting}. Most of them follow a generation-based paradigm, prompting an LLM to output a ranked list of candidate identifiers. However, this paradigm is costly due to autoregressive decoding, and it is sensitive to prompt design and instruction-following errors, which can lead to invalid outputs, missing candidates, or unstable rankings.

To avoid the brittleness and overhead of generation-based paradigm, an alternative line of work estimates relevance from internal LLM signals.~\citet{chenattention} propose In-Context Re-ranking (ICR), which feeds the documents and query into an LLM and compute document relevance scores by aggregating query-to-document attention across all attention heads. However, aggregating over all heads can mix informative signals with redundant or noisy patterns, which can degrade ranking quality. QRhead~\citep{zhang2025query} shows that a small set of heads can be sufficient.  Specifically, it constructs a head-detection dataset that contains queries, candidate documents, and a labeled gold document for each query. It then scores each head by its attention mass on the gold document, averages the scores across queries, and selects the top-$k$ heads (e.g., $k=16$) for re-ranking. However, this scoring rule does not penalize heads that also attend strongly to irrelevant documents. CoRehead~\citep{CoRehead} addresses this with a contrastive criterion that prefers heads that allocate more attention to the gold document while allocating less attention to irrelevant documents.

\begin{figure}[t]
    \centering
    \includegraphics[width=1.0\linewidth]{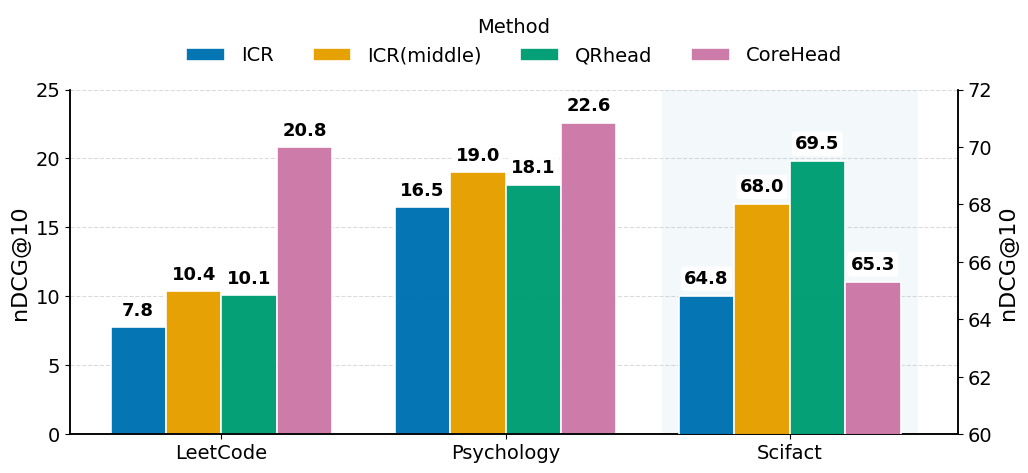}
    \caption{Performance comparison (nDCG@10) across three datasets for different attention-based re-ranking methods on Llama-3.2-1B. ICR(middle) denotes the ICR variant that directly selects all attention heads from the middle layers (7-12).}
    \label{fig:toy_example}
\end{figure}

Despite recent progress, head selection is often treated as a static decision: a single set of heads, identified on a head-detection dataset, is reused across queries and domains. Figure~\ref{fig:toy_example} shows that the impact of head selection varies substantially across datasets. First, specialized head selection does not consistently outperform a simple layer-based strategy, ICR (middle), which directly selects all heads from the middle layers. Second, the performance of head selection methods differs across datasets: CoreHead performs best on LeetCode and Psychology, whereas QRhead performs best on Scifact. These results indicate that head utility is strongly dataset-dependent, and that a static head selection rule does not generalize reliably.

To achieve better fine-grained re-ranking, we propose a query-dependent head routing method that dynamically selects attention heads based on query features, called \textbf{RouteHead}. Training such a router requires supervision that maps each query to an optimal head set, but the annotations are unavailable in practice. We therefore adopt a two-stage pipeline. First, we construct pseudo labels via offline search. The search uses forward selection with early stopping, followed by swap-based local refinement, and produces a multi-hot label for each query. Second, we train a router for query-dependent head selection. Specifically, the router assigns each attention head a learnable embedding and represents each query using an embedding extracted from frozen LLM hidden states. It then scores each query--head pair and applies an independent sigmoid to obtain activation probabilities. It is trained by minimizing binary cross-entropy against the multi-hot pseudo labels, together with a sparsity regularizer. At inference time, the router dynamically selects a small set of heads, and relevance scores are computed using only these heads. Our contributions are as follows:

1. We introduce a lightweight query-to-head router that enables real-time, query-dependent head selection for attention-based re-ranking with LLMs.

2. We propose an offline search procedure to build pseudo labels that approximate optimal query-to-head labels for router training.

3. We evaluate the proposed method against strong baselines across multiple LLM backbones on diverse benchmarks, and it consistently achieves the best performance.

\section{Preliminary and Problem Definition}
\label{sec:problem}

Given a query $q$ and a set of $N$ retrieved candidate documents $D=\{d_1,\dots,d_N\}$,  we feed the concatenated sequence $X=[d_1,d_2,\dots,d_N,q]$ into the LLM with $M=\{h_1, h_2,\dots,h_M\}$ attention heads.  For a document $d_i$, we define its relevance score under head $h_m$ as:
\begin{equation}
  s_{h_m}(d_{i},q) = \frac{1}{|\mathcal{I}_q|} \sum_{j \in \mathcal{I}_{d_i}} \sum_{z \in \mathcal{I}_q} a^{h_m}_{z\rightarrow j}
\end{equation}
where $\mathcal{I}_q$ and $\mathcal{I}_{d_i}$ are the index sets of query tokens and tokens in $d_i$, respectively, and $a^{h_m}_{z\rightarrow j}$ denotes the attention weight from the $z$-th query token to the $j$-th token of $d_i$ under head $h_m$.

\paragraph{Problem definition.}
Our goal is to learn a router $\phi$ that can route each query $q$ to an optimal head set $\phi(q)$ and aggregates the document relevance scores from the selected heads for re-ranking:
\begin{equation}
s(d_i,q)
=
\sum_{h_m \in \phi(q)} s_{h_m}(d_i,q)
\label{eq:set_agg}
\end{equation}

\section{Methodology}
\label{sec:method}
Our method follows a two-stage pipeline, which consists of an offline search procedure for pseudo-label construction and a training procedure for the router. We describe each stage in detail below.

\subsection{Offline Pseudo-Label Construction}
\label{sec:teacher}

\paragraph{Search space.}

Searching over all heads is impractical for modern LLMs (e.g., 1024 heads in Llama-3.1-8B), since it makes head-set search either too expensive or too crude to yield high-quality sets. We therefore restrict the search space. Specifically, we score each head by the nDCG@10 it achieves when used alone for re-ranking and retain only the top-$K$ heads (e.g., $K=64$) as a compact head pool $\mathcal{H}$ . This focuses the subsequent search on a smaller, higher-quality pool, improving both efficiency and the resulting head sets.

\paragraph{Search with an adaptive budget.}
For each query $q$, the goal is to identify a head set $\mathcal{S}(q)\subset \mathcal{H}$ to approximate $\phi(q)$ that maximizes ranking quality, measured by nDCG@10 and denoted by $\mathcal{L}_{\mathrm{rank}}(q,\mathcal{S})$. Although exhaustive search can yield the optimal solution, it is computationally infeasible in practice, requiring evaluation of $2^{K}$ subsets per query.  We therefore approximate the query-wise optimum with a two-phase search procedure. We set a maximum set size $P$ (e.g., $P=8$), but allow $|\mathcal{S}(q)|<P$ because the best performance can occur with fewer heads. In the first phase, we perform forward selection with early stopping. Starting from $\mathcal{S}_0=\emptyset$, at step $t\ge 1$ we add the head $h_m$ that maximizes the objective after inclusion,
\begin{equation}
h_t
=
\arg\max
\mathcal{L}_{\mathrm{rank}}\big(q, \mathcal{S}_{t-1}\cup\{h_m\}\big),
\qquad
\mathcal{S}_t=\mathcal{S}_{t-1}\cup\{h_t\}
\label{eq:forward_select}
\end{equation}
and stop at the first $T$ such that either $|\mathcal{S}_T|=P$ or no remaining head improves the objective. This stage produces an initial set $\mathcal{S}_T$ with $|\mathcal{S}_T|\le p$. The initial set can depend on the order in which the heads are considered. To reduce the dependency and explore alternative combinations of heads, in the second phase, the set is refined using swap-based local search. Let $\mathcal{S}^{0}=\mathcal{S}_{\mathrm{T}}$ and let $\mathcal{S}^{r}$ denote the current set at iteration $r$. The search considers one-swap moves that replace a selected head $h_u\in \mathcal{S}^{r}$ with an unselected head $h_v\in \mathcal{H} \setminus \mathcal{S}^{r}$, and selects the best swap by
\begin{equation}
(h_u,h_v)
=
\arg\max
\mathcal{L}_{\mathrm{rank}}\!\big(q,\; (\mathcal{S}^{r}\setminus\{h_u\})\cup\{h_v\}\big)
\label{eq:swap_argmax}
\end{equation}
Let $\mathcal{S}'=(\mathcal{S}^{r}\setminus\{h_u\})\cup\{h_v\}$. The swap is accepted if it improves the objective by at least a tolerance $\epsilon\ge 0$,
in which case the update $\mathcal{S}^{(r+1)}\leftarrow \mathcal{S}'$ is applied; otherwise, the procedure terminates and $\mathcal{S}(q)=\mathcal{S}^{r}$. The final head set defines a multi-hot label for query $q$:
\begin{equation}
y_{h_m}(q)=
\begin{cases}
1, & h_m \in \mathcal{S}(q),\\
0, & \text{otherwise}.
\end{cases}
\label{eq:multi_hot}
\end{equation}

\begin{table*}[!]
    \centering
        \renewcommand{\tabcolsep}{1.2mm}
        \renewcommand{\arraystretch}{1.0}
    \fontsize{8}{8}\selectfont
    % \footnotesize
    \begin{tabular}{lcccccccccccc}
    \toprule
& \bf NQ & \bf COVID & \bf NFCorpus & \bf FiQA & \bf Scifact & \bf Scidocs & \bf FEVER & \bf Climate & \bf DBPedia & \bf Robust04 & \bf News & \bf Avg \\
\midrule
\rowcolor{gray}BM25 \dag{} & 30.5 & 59.5 & 32.2 & 23.6 & 67.9 & 14.9 & 65.1 & 16.5 & 31.8 & 40.7 & 39.5 & 38.4 \\
$\text{RankGPT}_\text{Llama-3.1-8B}$ & 53.7 & 75.5 & 34.3 & 31.4 & 69.3 & 17.4 & 67.5 & 23.8 &  
42.9 & 47.8 & 46.2 & 46.3 \\
$\text{RankGPT}_\text{Qwen-2.5-7B}$ & 42.7 & 70.5 & 34.1 &  29.5 & 69.3 & 16.6 & 70.5 & 19.7 & 37.1 &  46.4 &  43.6 & 43.6 \\
\midrule
& \multicolumn{11}{c}{\textit{\textbf{Base LLM: Llama-3.2-1B-Instruct}}}  \\
ICR & 35.6& 57.2& 28.8& 22.3& 64.8& 13.7& 75.7& 16.8& 25.9& 36.0& 37.8&37.7 \\
QRhead &  41.4 &  67.3 &  31.6 & 25.1 &  69.5 &  15.5 & 77.4 &  23.3 & 31.8 & 42.2 & 42.2 &  42.5 \\
Corehead &  35.6 & 56.9& 29.1& 24.8& 65.3& 14.3& 70.6& 17.4& 21.8& 35.1& 34.9& 36.9 \\
RouteHead & 42.4 & 67.7 & 32.0 & 25.9 & 70.1 & 15.8 & 78.0 & 23.4 & 32.4 & 42.8 & 42.3 & 43.0 \\

\midrule
& \multicolumn{11}{c}{\textit{\textbf{Base LLM: Llama-3.2-3B-Instruct}}}  \\
ICR & 49.3 & 72.5 & 33.8 & 31.7 & 73.3 & 17.4 & 82.7 & 24.2 & 34.7 & 47.1 & 44.6 & 46.5 \\
QRhead &  54.9 &  77.3 &  35.1 & 35.0 &  74.7 &  18.3 &  83.6 &  24.5 & 36.0 & 49.7 & 45.2 &  48.6 \\
Corehead  &51.4& 77.3& 34.2& 35.0& 73.8& 18.4& 82.0& 23.2& 32.0 & 48.7 &43.2 &47.2\\
RouteHead & 56.3 & \textbf{78.6} & 35.6 & 36.4 & 75.1 & 18.9 & 84.4 & 25.3 & 37.6 & 50.9 & 46.4 & 49.6 \\

\midrule

& \multicolumn{11}{c}{\textit{\textbf{Base LLM: Llama-3.1-8B-Instruct}}}  \\
ICR & 54.0 & 73.3 & 34.8 & 35.6 & 75.5 & 19.0 & 85.8 & 24.8 & 36.9 & 49.0 & 44.5 & 48.5 \\
QRhead & 58.6 & 77.5 & 35.3 & 39.1 & 76.2 &  19.4 & 85.3 & 23.8 & 37.2 &  51.4 &  46.1 & 50.0 \\
Corehead &54.0 & 75.6 & 34.1 & 39.4& 74.7 & 18.6 & 81.5 & 21.6 & 32.6 &  48.9 &  42.7& 47.5 \\
RouteHead & \textbf{59.8} & 78.2 & \textbf{36.0} & \textbf{39.6} & \textbf{76.7} & \textbf{19.9} & 85.8 & 24.0 & 37.6 & \textbf{51.6} & \textbf{46.8} & \textbf{50.5} \\

\midrule
& \multicolumn{11}{c}{\textit{\textbf{Base LLM: Qwen-2.5-7B-Instruct}}} \\
ICR & 43.1 & 66.1 & 32.7 & 27.0 & 71.1 & 16.4 & 79.2 & 19.6 & 35.3 & 43.0 & 40.0 & 43.0 \\
QRhead & 49.9 & 67.7 & 33.1 & 29.2 & 71.0 & 15.3 &  80.7 &  20.1 & 35.7 & 43.7 & 39.8 &  44.2 \\
Corehead &50.1 &67.7& 32.5& 30.6& 70.0&14.5 &79.8 &19.3 & 32.4& 45.0&42.1&44.0\\
RouteHead & 51.4 & 68.0 & 33.4 & 31.2 & 71.2 & 16.0 & 81.2 & 20.4 & 36.3 & 44.5 & 40.2 & 44.9 \\
\midrule
& \multicolumn{11}{c}{\textit{Traditional re-rankers}} & \\
% \cmidrule{0-12}
\rowcolor{gray}Contriever & 44.6 & 67.5 & 32.8 & 28.4 & 67.1 & 18.9 & 64.2 & 28.0 & 39.5 & 45.7 & 41.7 & 43.5 \\
\rowcolor{gray}GTR-T5-base & 51.4 & 74.8 & 32.5 & 34.7 & 62.1 & 15.8 & 72.9 & 26.8 & 37.1 & 46.1 & 42.8 & 45.2 \\
\rowcolor{gray}BGE-Reranker-base & 55.2 & 66.4 & 31.0 & 31.7 & 70.8 & 15.7 & \textbf{88.6} & \textbf{36.5} & 42.5 & 39.9 & 37.0 & 46.8 \\
\rowcolor{gray}MSMARCO-MiniLM & 55.8 & 74.3 & 35.2 & 35.1 & 68.5 & 17.5 & 80.4 & 25.5 & \textbf{45.3} & 47.9 & 43.0 & 48.0 \\
\bottomrule
    \end{tabular}
    \caption{Performance comparison (nDCG@10) on BEIR.  \dag{} represent initial retrieval. Best results are highlighted in \textbf{Bold}.}
    % \textbf{Bold} indicates the best results
    \label{tab:beir_main}

\end{table*}

\subsection{Router Training}
\label{sec:router}

\paragraph{Router Architecture Design.} Drawing inspiration from matrix factorization models in recommendation systems, we route query to heads using low-dimensional embeddings. Since head indices are discrete and non-differentiable, we assign each attention head $h_m$ a learnable embedding $e_{h_m}\in\mathbb{R}^{d_h}$ to represent head identity for routing. This continuous parameterization allows the router to organize heads into functional regions in a shared embedding space and to share selection patterns across heads with similar behavior. For query representation, we extract a query embedding $e_q\in\mathbb{R}^{d_q}$ from the frozen LLM hidden states using mean pooling over token representations. Given $(e_q, e_{h_m})$, the router computes a query-dependent activation score for each head via a bilinear-style interaction:
\begin{equation}
\alpha_{h_m}(q)
=
W^\top_2 \Big(
e_h \odot (W^\top_1 e_q + b)
\Big)
\label{eq:router_score}
\end{equation}
where $\odot$ denotes the Hadamard product, $W^\top_1 \in \mathbb{R}^{d_q \times d_h} $ and $b\in \mathbb{R}^{d_h} $ are parameters of a projection layer that align the dimension of $e_q$ with $e_{h_m}$, and $W^\top_2 \in \mathbb{R}^{d_h} $ is the weight vector of a linear layer that produces a scalar score. The score is then converted into an independent activation probability using the sigmoid function $\sigma$:
\begin{equation}
p_{h_m}(q)=\sigma\big(\alpha_{h_m}(q)\big)
\label{eq:router_prob}
\end{equation}

\paragraph{Training Objective.} Let $y_{h_m}(q)\in\{0,1\}$ denote the set indicator of whether head $h_m$ is selected for query $q$ by the offline labeling stage. We train the router by minimizing a set-level binary cross-entropy loss:
\begin{equation}
\mathcal{L}_{\mathrm{route}}
=
\sum_{m=1}^{K}
\mathrm{BCE}\big(p_{h_m}(q),\, y_{h_m}(q)\big)
\label{eq:route_loss}
\end{equation}

Because the router outputs independent, non-normalized activation probabilities across heads, an explicit constraint is required to control the number of activated heads. To mitigate over-activation caused by noisy or redundant pseudo labels, an $\ell_1$-style penalty is added on the activation probabilities:
\begin{equation}
\mathcal{L}_{\mathrm{sparse}}
=
\lambda \sum_{m=1}^{K} p_{h_m}(q)
\label{eq:sparse}
\end{equation}
This regularizer discourages uniformly high activation probabilities across heads. Thus the total objective is
\begin{equation}
\mathcal{L}
=
\mathcal{L}_{\mathrm{route}}+\mathcal{L}_{\mathrm{sparse}}
\label{eq:total_loss}
\end{equation}

% \subsection{Inference}
% \label{sec:inference}

% Given a query $q$, the router computes $p_h(q)$ for all heads and forms a head set
% \begin{equation}
% S(q)=\{\,h \mid p_h(q)>\tau\,\},
% \label{eq:threshold}
% \end{equation}
% where $\tau$ is a threshold. Alternatively, one can select the top-$K$ heads by $p_h(q)$ to match a fixed compute budget. The final document score is computed using Eq.~\ref{eq:set_agg} and used to rank the candidates.

 \begin{table*}[t] 
  \centering
        \renewcommand{\tabcolsep}{1.2mm}
        \renewcommand{\arraystretch}{1.0}
    \fontsize{8}{8}\selectfont
  \begin{tabular}{ccccccccccccccc} 
    \toprule 
    \multirow{2}{*}{\textbf{Model}} & \multicolumn{7}{c}{\textbf{StackExchange}} & \multicolumn{2}{c}{\textbf{Coding}} & \multicolumn{3}{c}{\textbf{Theorem-based}} &  \multirow{2}{*}{\textbf{Avg.}}\\ 
    \cmidrule(lr){2-8} \cmidrule(lr){9-10} \cmidrule(lr){11-13} 
   
     & Bio. & Earth. & Econ. & Psy. & Rob. & Stack. & Sus. & Pony & LC. & AoPS & TheoT. & TheoQ. & \\
    \midrule
    \rowcolor{gray}BM25\dag{}  & 18.2 & 27.9 & 16.4 & 13.4 & 10.9 & 16.3 & 16.1 & 4.3 & 24.7 & 6.5 & 2.1 & 7.3 & 13.7 \\ 
    $\text{RankGPT}_\text{GPT4}$ & 33.8 & 34.2 & 16.7 & 27.0 & 22.3 & 27.7 & 11.1 & 15.6 & 3.4 & 1.2 & 8.6 & 0.2 & 16.8 \\ 
    $\text{RankGPT}_\text{Qwen3-32B}$ & 24.9 & 29.4 & 20.9 & 25.7 & 18.3 & 16.0 & 23.2 & 7.6 & 27.6 & 7.8 & 8.9 & 8.4 & 18.2 \\
    $\text{RankGPT}_\text{Qwen3-235B-A22B}$ & 26.4 & 26.7 & 22.1 & 26.3 & 18.8& 17.0 & 24.9 & 8.2 & 27.2 & 7.7 & 11.7 & 8.6 & 18.8 \\
     \midrule

   \multicolumn{14}{c}{\textit{\textbf{Base LLM: Llama-3.2-1B-Instruct}}}  \\
   ICR &26.2 &28.9&	11.8&16.5&8.7&6.1	&11.5&	2.7&	7.8&	6.0&	2.6& 4.0&11.1\\
   QRhead &28.4 &32.7&15.3&18.1 &11.0 &6.8 &12.0&3.0&10.1&6.3&3.4&5.0 & 12.6\\
   Corehead &28.6 & 38.2& 19.2& 22.6& 13.8 &11.8& 16.9 &3.5&20.8 &10.7 &4.6& 10.1 &16.7\\
   RouteHead & 29.6 & 39.0 & 20.0 & 23.5 & 14.6 & 11.3 & 17.8 & 3.3 & 20.3 & 11.4 & 5.2 & 10.0 & 17.2\\

   \midrule

   \multicolumn{14}{c}{\textit{\textbf{Base LLM: Llama-3.2-3B-Instruct}}}  \\
   ICR &33.4& 37.7& 18.5& 21.2 &13.8 &10.0 & 17.4& 2.4& 13.3& 7.4& 2.9& 7.9 & 15.5\\
   QRhead &31.5& 40.5& 22.3& 27.9 &16.5 &12.7 & 20.0& 2.8& 18.8& 7.8&5.5& 9.7& 18.0\\
   Corehead &30.7& 40.5& 22.7& 27.2& 16.4& 14.6& 20.1& 3.4& 22.4& 9.5& 5.6& 10.8 &18.6\\
 
   RouteHead &\textbf{33.9}&\textbf{41.2}&\textbf{23.1}&\textbf{28.8}&17.3&13.5&20.8&3.0&19.5&8.4&6.2&10.2&18.9\\
    
    \midrule
    \multicolumn{14}{c}{\textit{\textbf{Base LLM: Llama-3.1-8B-Instruct}}}  \\
   ICR &31.1& 33.3& 18.2& 20.5& 15.7& 13.5& 17.8& 2.4& 16.8& 8.3&5.2&11.3&16.2\\
   QRhead &30.8 &35.5 &20.2 & 25.1& 18.9&15.2& 20.3& 3.6& 20.9& 9.8 &5.5 &11.8& 18.1 \\
   Corehead &31.2& 37.9& 22.3& 26.9& 18.2& 16.7& 22.1& \textbf{3.7}& \textbf{24.6}& 11.0& 6.7& \textbf{12.8}&19.5\\
   RouteHead & 32.0 & 38.4 & 22.8 & 27.6 & \textbf{18.8} & \textbf{16.2} & \textbf{22.8} & 3.5 & 24.0 & \textbf{11.4} & \textbf{7.3} & 12.5 & \textbf{19.8}\\
   \midrule

   \multicolumn{14}{c}{\textit{\textbf{Base LLM: Qwen-2.5-7B-Instruct}}}  \\
   ICR &27.6& 32.6& 14.5& 19.2& 14.0& 9.4& 16.5& 2.0& 15.7& 7.2& 3.3& 6.7 &14.0\\
   QRhead  &26.8 &34.7 & 15.1& 21.6& 13.9& 8.9& 18.5& 3.3& 16.8& 6.3& 4.3& 8.0  &14.8\\
   Corehead  &28.3 &35.1& 16.7& 23.8& 15.1& 8.7& 19.0& 4.9& 18.6& 7.6& 4.7& 8.8&15.9\\
   RouteHead &29.3&36.5&17.3&23.8&15.5&10.4&20.0&4.1&18.2&7.3&5.5&9.2&16.5\\

    \bottomrule % Bottom line
  \end{tabular}%
\caption{Performance comparison (nDCG@10) on BRIGHT . \dag{} represent initial retrieval. Best results are highlighted in \textbf{Bold}.}

  \label{tab:bright}
  
  \end{table*}

\section{Experiments}
\label{sec:main_exp}

\subsection{Experimental Setup}

\paragraph{Datasets.} We evaluate on two benchmarks. The first is BEIR~\citep{thakur2021beir} with eleven sub-datasets across diverse domains: NQ~\citep{NQ}, COVID~\citep{voorhees2021treccovid}, NFCorpus~\citep{boteva2016nfcorpus}, FiQA~\citep{maia2018fiqa}, SciFact~\citep{wadden-etal-2020-scifact}, SciDocs~\citep{cohan2020scidocs}, FEVER~\citep{thorne2018fever}, Climate~\citep{diggelmann2020cfever}, DBPedia~\citep{hasibi2017dbpedia}, Robust04~\citep{jeronymo2022mrobust04} and News. The second is BRIGHT~\citep{su2025bright}, a reasoning-intensive retrieval benchmark with 1{,}385 real-world queries from multiple domains (e.g., StackExchange, LeetCode, and math competitions).

\paragraph{Baselines.} 
We compare our approach with baselines from two paradigms: the generation-based method \textbf{RankGPT}~\cite{sun2023RankGPT} and the attention-based methods \textbf{ICR}~\citep{chenattention}, \textbf{QRhead}~\cite{zhang2025query}, and \textbf{CoRehead}~\citep{CoRehead}. We also include several traditional re-rankers: Contriever~\citep{Contriever}, GTR-T5-base~\citep{GTR-T5-Base}, BGE-Reranker-base~\citep{BGE-reranker} and  MSMARCO-MiniLM~\citep{reimers-gurevych-2019-sentence}.

\paragraph{Base LLMs.} We evaluate instruction-tuned LLMs from two families and multiple scales, including Llama-3.2-1B, Llama-3.2-3B and Llama-3.1-8B from the Llama series, as well as Qwen2.5-7B from the Qwen series.  We also integrate the open-source reasoning models Qwen3-32B and Qwen3-235B-A22B into RankGPT to evaluate performance on the reasoning-intensive BRIGHT benchmark.

\paragraph{Setting.} We re-rank the top-200 BM25 candidates and randomly sample 512 queries for evaluation on BEIR. On BRIGHT, we re-rank the top-100  BM25 candidates and evaluate on the full query set. RankGPT uses bubble-sort with a sliding window of size 20 and stride 10, while others re-rank the full candidate list.

\subsection{Experiment Results}
We report results on BEIR in Table~\ref{tab:beir_main} and on BRIGHT in Table~\ref{tab:bright}. Across both benchmarks and all base LLMs, RouteHead achieves the best average performance among attention-based methods, and performance increases monotonically within each model family as the backbone size increases. This suggests that our method provides complementary gains on top of stronger LLM backbones. Specifically, on BEIR, RouteHead consistently outperforms the strongest static head baseline (QRhead), indicating that a single globally selected head set does not transfer reliably across heterogeneous IR collections. Under the same evaluation setting, RouteHead  with Llama-3.2-3B reaches an average of 49.6, exceeding the strongest traditional re-ranker, MSMARCO-MiniLM (48.0). It also outperforms the generation-based method RankGPT while using a smaller backbone, surpassing $\text{RankGPT}_\text{Llama-3.1-8B}$ (46.3) and $\text{RankGPT}_\text{Qwen-2.5-7B}$ (43.6) with Llama-3.2-3B.  Although RouteHead underperforms traditional re-rankers on Climate and DBPedia, this gap is not specific to RouteHead: the LLM-based re-ranking methods are generally weaker than strong end-to-end cross-encoders on these two datasets, which emphasize domain terminology and entity-centric matching. 

RouteHead remains competitive even under reasoning-intensive benchmark BRIGHT. With Llama-3.2-3B, it reaches 18.9, exceeding the RankGPT result with Qwen3-235B-A22B (18.8). Notably,  RouteHead is often below CoreHead on coding- and math-oriented domains, whereas it is consistently strong on the StackExchange subsets and often improves multiple domains. One plausible explanation is a training--evaluation distribution mismatch. The router is trained on MSMARCO, whose queries primarily reflect general web information needs and may provide limited coverage of code- and math-oriented questions. Consequently, the learned routing patterns may be less aligned with the highly structured signals required by coding- and theorem-style queries. This observation motivates improving the coverage of router training data, for example by incorporating more code and math queries or by using domain-balanced sampling during pseudo-label construction and router training. Together, these results show that RouteHead delivers strong and consistent improvements within the attention-based re-ranking, , while remaining competitive against stronger traditional and generation-based baselines. 

% Because only a lightweight router is trained on top of a frozen backbone, RouteHead also offers a practical quality--efficiency trade-off.

\section{Related Work}

Our work is related to both zero-shot re-ranking with LLMs and the attention mechanism. Zero-shot re-ranking with LLMs is typically formulated in point-wise~\citep{sachan2022improving, liang2022holistic,mo2023convgqr}, pair-wise~\citep{qin2024large,meng2025query,xu2025survey}, or list-wise~\citep{ma2023zero, sun2023RankGPT, chenattention, chen2023extending, jin2024llm, fu2024data} manners. However, most of them adopt a generation-based paradigm, which incurs high cost and latency due to autoregressive decoding~\citep{zhang2026context} and is sensitive to prompt design and instruction-following errors. Attention is a central mechanism in transformer LLMs and provides signals that are informative for retrieval~\citep{izacard2021distilling,reattn,mo2026opendecoder,zhang2026starpo}. \citet{chenattention} treat attention as an implicit relevance signal and compute document relevance scores for re-ranking by aggregating query-to-document attention weights across all heads. Earlier mechanistic interpretability studies suggest that only a subset of attention heads is functionally critical~\citep{michel2019sixteen, voita2019analyzing,zhang2024blind,wu2025retrieval,su2026parametric,zhang2025ratt,zhang2026preferenceheads}. Motivated by these findings, recent work proposes heuristic rules to identify a small set of important heads for attention-based re-ranking~\citep{zhang2025query, CoRehead}. However, these methods typically select a fixed head set, which can be suboptimal since the useful heads can differ across queries. This limitation motivated our exploration of real-time, query-dependent head selection.

\section{Conclusion}

We introduce a lightweight router that learns to route queries to attention heads for attention-based re-ranking with LLMs, where relevance scores are computed by aggregating attention signals from the selected heads. The router is trained using pseudo labels constructed by an offline search procedure. Extensive experiments demonstrate the effectiveness of our method. Future work can explore improving the quality and consistency of pseudo labels, enriching the router inputs with additional information (e.g., query type and length).

\bibliographystyle{ACM-Reference-Format}
\bibliography{sample-base}

\appendix

\end{document}